\documentclass[aps,prl,showpacs,twocolumn,groupedaddress]{revtex4-1}
\usepackage{graphicx}
\usepackage{mathtools}
\usepackage{amssymb}
\usepackage{color}
\usepackage{float}
\usepackage{gensymb}

\DeclareFontFamily{U}{mathb}{\hyphenchar\font45}
\DeclareFontShape{U}{mathb}{m}{n}{
      <5> <6> <7> <8> <9> <10> gen * mathb
      <10.95> mathb10 <12> <14.4> <17.28> <20.74> <24.88> mathb12
}{}
\DeclareSymbolFont{mathb}{U}{mathb}{m}{n}
\DeclareMathSymbol{\rightsquigarrow}{3}{mathb}{"F9}

\begin{document}

\title{Post-firewall paradoxes}

\author{Samuel L.\ Braunstein}
\email[]{sam.braunstein@york.ac.uk}
\homepage[]{http://www.cs.york.ac.uk/~schmuel}
\affiliation{Computer Science, University of York, York YO10 5GH, 
United Kingdom}

\author{Stefano Pirandola}
\affiliation{Computer Science, University of York, York YO10 5GH, 
United Kingdom}

\date{\today}

\begin{abstract}

The pre\"eminent view that evaporating black holes should
simply be smaller black holes has been challenged by the firewall
paradox. In particular, this paradox suggests that something different
occurs once a black hole has evaporated to one-half its original surface
area. Here we derive variations of the firewall paradox by tracking the
thermodynamic entropy within a black hole across its entire lifetime.
Our approach sweeps away many unnecessary assumptions, allowing us to
demonstrate a paradox exists even {\em after} its initial onset (when
conventional assumptions render earlier analyses invalid). Our results
suggest that not only is the formation of a firewall the most natural
resolution, but provides a mechanism for it. Finally, although firewalls
cannot have evolved for modest-sized black holes, within the age of the
universe, we speculate on the implications if they were ever
unambiguously observed.
\end{abstract}


\maketitle

The fundamental physics of black holes has been an enduring mystery
\cite{Hawking76}. Great progress has been recently made with the
discovery of the firewall paradox for black holes, which suggests the
existence of a manifestly strong phenomenon, the {\it firewall\/}
\cite{AMPS} or {\it energetic curtain}\cite{Braunstein13} as it was
originally dubbed in 2009. Loosely, the paradox constructs a
contradiction between the correspondence relating real to classical
black holes, their thermodynamic behavior and quantum unitarity.
However, finding the minimal assumptions has remained an open question.

Here, we provide a streamlined firewall paradox with many unnecessary
assumptions removed. Notably, no measurement and consequently no
decoding of the Hawking radiation is needed (making its complexity
\cite{Hayden} irrelevant), nor do we rely on any specific radiation
process like pair creation or tunneling. This last point deserves
especial comment: First, if firewalls were real, then `nice time slices'
through a black hole's spacetime \cite{Lowe95} and all mechanisms
associated with them could no longer be trusted. Second, black hole
evaporation is strongly believed to be non-local (e.g., via the
holographic principle\cite{Hooft,Susskind95}). However, pair creation in
particular comes from {\em local} quantum field theory \cite{Hawking76}.
These considerations render earlier derivations of the firewall paradox
\cite{AMPS} problematic. Next, we derive a new paradox which
independently explores the role of non-local physics across the horizon.
We find that there is an extra ingredient left out of conventional
holographic approaches \cite{Maldacena,Hooft,Susskind95}. Combined,
these two paradoxes provide insight into black hole physics across their
entire lifetime. Evidence supporting our assumptions and variations to
them are given in the discussion.

Before proceeding, let us review an important tool: The quantum mutual
information, $S(X:Y)\equiv S(X)+S(Y)-S(X,Y)$, provides a measure of
correlations between a pair of systems $X$ and $Y$. Here the von Neumann
entropy, denoted $S(X)$, gives the thermodynamic entropy for an {\it
isolated\/} system\cite{vN}, $X$.  As with some earlier works studying
black hole evaporation \cite{Mathur09,AMPS} we shall rely on the
property of strong subadditivity \cite{Wilde}, which for any extra
system $W$, states that $S(W,X\!:\!Y)\ge S(X\!:\!Y)$.

We will leverage this tool to identify contributions to thermodynamic
entropy in even very large quantum systems.  Suppose $X$ and $Y$ are
{\it remotely\/} separated and no longer interact. If the correlations
between these systems correspond to maximally entangled subspaces of $X$
and $Y$, then they make a distinct contribution \cite{fn3} of
$\frac{1}{2}S(X\!:\!Y)$ to the thermodynamic entropy in $X$ (and in
$Y$); an entropy that is observable by even highly delocalized
detectors. This is in contrast to local correlations due, for example,
to entropy of entanglement in the vacuum state across a boundary
\cite{Eisert10}. In that case, correlations are localized to a narrow
layer at the boundary and unless one's detectors are localized on scales
comparable to the cutoff (presumably Planckian) such entropy is
unobservable.

{\it Non-exotic atmosphere}: To reconcile gravity with quantum
mechanics, it is generally assumed that there exists a correspondence
between the physical characteristics of a real (i.e., quantum
mechanical) black hole and its theoretical classical counterpart. The
tightest correspondence \cite{Hawking76} assumes that black holes
evaporate into vacuum (as seen by an infalling observer). Here, our two
theorems shall make much weaker assumptions than that the quantum field
into which the black hole evaporates is in (or anywhere near to) a
specific quantum state. 

In part, we achieve this by focusing on the gross thermodynamic
properties of the neighborhood, $N$, external to and surrounding a black
hole that reaches out far enough to encompass any process by which
radiation is produced. Now, recall 't Hooft's entropic bound
\cite{Hooft,Hsu08}, which shows that if one excludes configurations of
ordinary matter that will inevitably undergo gravitational collapse, one
finds ${\cal A}^{3/4}\ge S_{\text{matter}}$, where ${\cal A}$ is the
surface area of the region containing that matter (in Planck units) and
$S_{\text{matter}}$ is the thermodynamic entropy of the matter (with
Boltzmann's constant set to unity).

Suppose that the external neighborhood, at some specific epoch, has a
surface area $\mu$ times that of the black hole's horizon, so
${\cal A}_N=4 \mu\, S_{\text{BH}}$, where $S_{\text{BH}}$ denotes the black
hole's Bekenstein Hawking entropy. Then to satisfy 't Hooft's bound, the
thermodynamic entropy of the external neighborhood,
$S_{\text{therm}}^N$, must be bounded by
\begin{equation}
{\cal A}_N^{3/4}
\equiv \biggl[ 2\sqrt{2}\,
\Bigl(\frac{\mu^3}{S_{\text{BH}}}\Bigr)^{\!1/4}\biggr] S_{\text{BH}}
\ge S_{\text{therm}}^N.
\label{Correspondence}
\end{equation}
For large black holes, the prefactor in square brackets is much much
less than unity even for extremely large neighborhoods, e.g., $\mu
=10^4$. If this bound fails, the external neighborhood, $N$, {\it
must\/} consist of some exotic matter, such as an atmosphere of
microscopic black holes.

\noindent {\bf Theorem 1}: A contradiction exists between: 1.a)
completely unitarily evaporating black holes, 1.b) a freely falling
observer notices nothing special until they pass well within a large
black hole's horizon, and 1.c) the black hole interior Hilbert space
dimensionality may be well approximated as the exponential of the
Bekenstein Hawking entropy.

\begin{figure}[ht]
\vskip -0.1truein
\includegraphics[width=38mm]{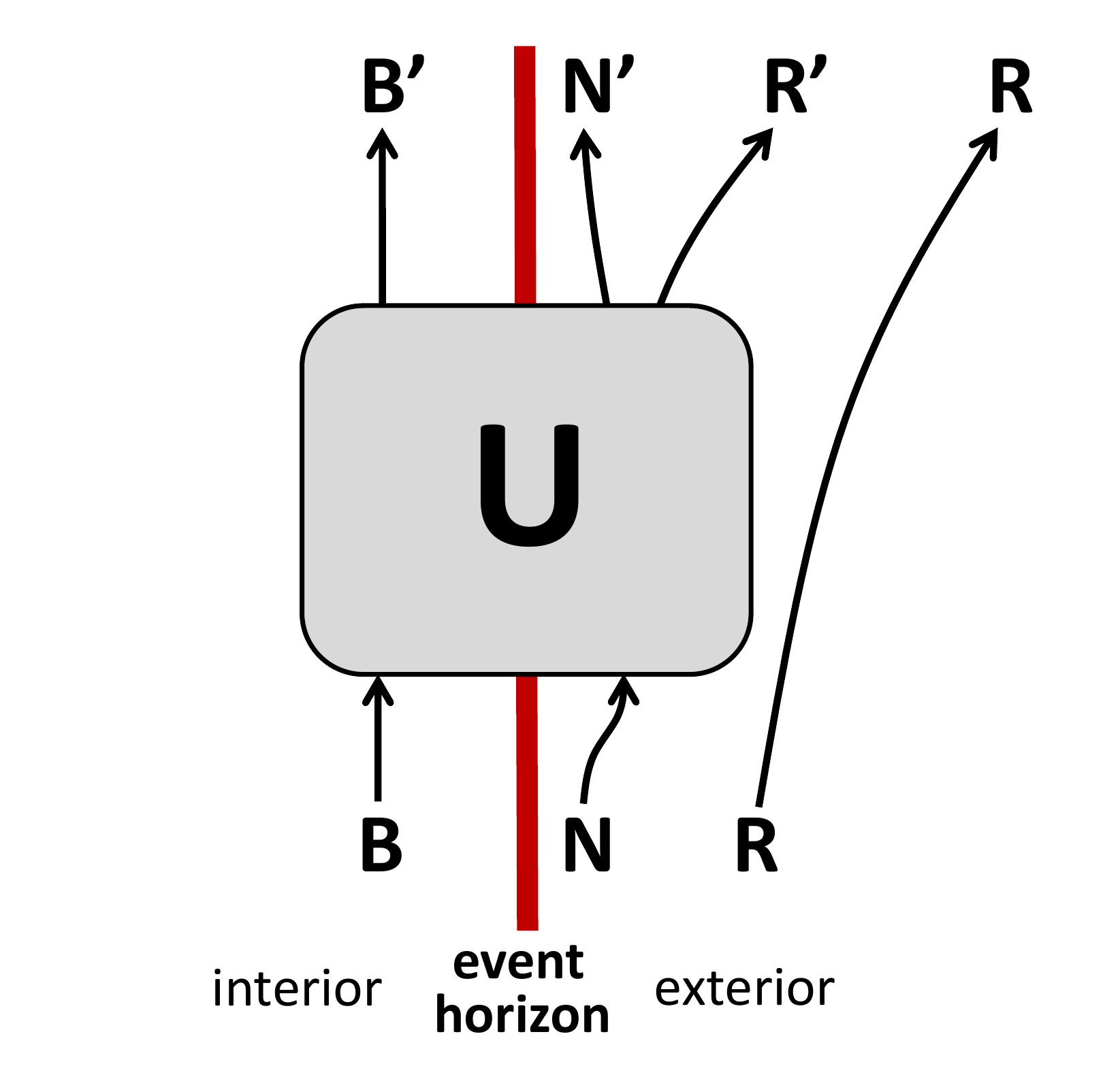}
\vskip -0.17truein
  \caption{
Schematic generation of radiation $R'$ during some epoch by an arbitrary
unitary process $U$. Here, $N$ and $N'$ are the degrees-of-freedom in
the black hole's exterior neighborhood prior and posterior to the
unitary operation, respectively; $B$ and $B'$ label interior
degrees-of-freedom; and $R$ denotes radiation from an earlier epoch.
[Note, the initial joint quantum state of $(B,N,R)$ is arbitrary.]
  \label{generic_Hawking_process}}
\end{figure}

\noindent {\bf Proof}: Assumption 1.a has two ingredients (i and ii):
That black hole evaporation is unitary and that it is complete
leaving behind no remnant. We start by unraveling the implications of
these two ingredients separately:

{\em Unitary evaporation} 1.a(i): Consider the unitary generation of
radiation from a black hole by an arbitrary process,
Fig.~\ref{generic_Hawking_process}. We associate this process with
some specific black hole and presume that to an excellent approximation
radiation is not produced further out beyond $N$.

\vskip -0.13truein

Applying strong subadditivity to Fig.~\ref{generic_Hawking_process}
trivially yields $S(B',N',R'\!:\!R)\ge S(R'\!:\!R)$. Further, as entropy
is invariant under unitary transformations
$S(B',N',R'\!:\!R)=S(B,N\!:\!R)$ and we obtain
\begin{equation}
S(B,N:R) \ge S(R':R)\label{paradox1}.
\end{equation}
This invariance has allowed us eliminate $B'$ and $N'$ from
Eq.~(\ref{paradox1}), thus allowing us to work with quantities $B$ and
$N$ which may be associated with a still large black hole.

{\it Complete evaporation} 1.a(ii): Consider a black hole created by
collapsing matter initially in a pure quantum state (the 
Appendices consider more general scenarios). After the black hole has
completely evaporated away the net radiation, $(R,R')$, should also be in a 
pure quantum state to preserve unitarity. Thus one might expect that
correlations would exist between the early and late epoch radiation, $R$
and $R'$, respectively.

In fact, the study of random unitary operations allows us to say much
more: Since the Hilbert space dimensionalities involved are so huge
Levy's lemma guarantees a generic behavior for entropy in all but a set
of measure zero of evaporation mechanisms \cite{Braunstein13,fnPir}. In
particular \cite{Page93,Braunstein13}, the entropy of the radiation
grows monotonically (at almost exactly the maximal rate of one bit's
worth of entropy per qubit of radiation emitted \cite{bit}) until the
{\it Page time\/} (when the black hole's area has halved). From the Page
time (PT) onward the overall entropy in the radiation decreases at the
same rate, reaching zero when evaporation is complete
\cite{Page93,Braunstein13}. Thus, $S(R)=S(R')=\min(\log_e|R|,\log_e|R'|)$.

Now recall that the Bekenstein Hawking entropy quantifies the thermal
entropy in a black hole from the first law of black hole thermodynamics.
This presumes, however, that the radiation lacks correlations. Thus by
ignoring correlations in the above results we may identify \cite{Manas}
the Bekenstein Hawking entropy of the {\em initial} black hole as
$S_{\text{BH}}= \log_e|R|+\log_e|R'|$, where $|X|={\text{dim}}(X)$.
Similarly, after evaporating radiation $R$, the remaining black hole has
Bekenstein Hawking entropy $\log_e|R'|$. Labeling the pre- and post-PT
as $|R| < |R'|$ and $|R| > |R'|$, respectively, then gives
\begin{eqnarray}
{\textstyle \frac{1}{2}}\,S(R':R)&=&
\min(\log_e|R|,\log_e|R'|)\nonumber \\
&=&\left\{
\begin{array}{lr}
S_{\text{BH}}-\log_e |R'|, & \text{pre-PT},\\
\log_e |R'|, & \text{post-PT}.
\end{array}
\right.
\label{RpR}
\end{eqnarray}

For black holes formed by matter in a pure quantum state, the (global)
state of $(B,N,R)$ may also be treated as pure implying $S(B,N\!:\!R)=
S(B\!:\!R)+S(N\!:\!R)$. Because it will be sufficient to consider only
really huge violations of assumption 1.b, we can decompose it into
external and internal constraints (i and ii).

{\it External free-fall equanimity} 1.b(i): An exceedingly weak
assumption is that a freely-falling observer notices no exotic matter at
least down to the horizon. Thus, from Eq.~(\ref{Correspondence}), 
the radiation-correlated contribution to the neighborhood's entropy
$\frac{1}{2}S(N\!:\!R)\le S_{\text{therm}}^N$ must be negligible.
Combining this with Eq.~(\ref{paradox1}) yields
\begin{equation}
{\textstyle \frac{1}{2}}\,
S(B:R)\gtrsim
{\textstyle \frac{1}{2}}\,
S(R':R).
\label{BR}
\end{equation}

{\it Finite interior Hilbert space} 1.c: In order to ensure that the
black hole interior (within the stretched horizon) can contain no more
physical entropy than can eventually evaporate away, it is assumed that
the interior has a Hilbert space dimensionality that is well
approximated by the exponential of the Bekenstein Hawking entropy. This
implies that $\log_e|B|\simeq \log_e|R'|$ during a black hole's
evaporation and hence from Eqs.~(\ref{RpR}) and~(\ref{BR}) we have
\begin{equation}
\!\log_e|B|\ge {\textstyle \frac{1}{2}}
S(B\!:\!R)\gtrsim\!
\left\{
\begin{array}{lr}
S_{\text{BH}}-\log_e|B|, & \!\!\text{pre-PT},{\!\!}\\
\!\log_e|B|, & \!\!\text{post-PT}.{\!\!}
\end{array}\!\!
\right.
\label{T1}
\end{equation}

So far there is no contradiction. However, Eq.~(\ref{T1}) tells us that
the black hole interior initially accumulates thermodynamic entropy at
the rate of (at least) one bit per qubit radiated. From the PT, the
shrinking interior (everything within the stretched horizon) is filled
with half of a maximally entangled state,
$\frac{1}{2}S(B\!:\!R)\simeq \log_e|B|$, corresponding to an
infinite-temperature thermal state.

{\it Internal free-fall equanimity} 1.b(ii): A contradiction ensues if
we assume that an infalling observer notices nothing (or at worst a low
entropy state) well within the horizon. Indeed, after the PT there is
nowhere left within the black hole (within the stretched horizon) where
an observer can exist without intimate contact with the
infinite-temperature interior.
{$~$}\hfill
\rule{2mm}{2mm}
\vskip 0.03truein

This result follows straightforwardly and differs from earlier arguments
\cite{AMPS} which invoke a likely impossible capacity for decoding
Hawking radiation \cite{Hayden}. In the Appendices we
extend our analysis to explicitly account for the negligible
thermodynamic entropy in $N$ and to exclude the physics of Planck-scale
black holes.

{\it The AdS/CFT correspondence crisis}: The strongest contender for a
fully unitary theory of black hole evaporation involves the AdS/CFT
correspondence \cite{Maldacena}, which formalizes the holographic
principle \cite{Hooft,Susskind95}. Unfortunately, this theory gives no
hint of a firewall nor that anything unusual might happen within the
horizon from the PT onward. The conventional hope is that this
discrepancy will be resolved once the ``dictionary'' relating the CFT
and black hole interior states is better understood.

Despite this hope, this discrepancy has precipitated such a crisis for
the AdS/CFT and holographic approaches that their creators have resorted
to a radical \cite{Maldacena13} (many in the community call it an absurd
\cite{Susskind14}) solution. In particular, it is proposed that all
maximal entanglement (denoted EPR) is associated with Einstein Rosen
(ER) bridges (a kind of non-traversable wormhole) \cite{Maldacena13}; it
is claimed  \cite{Maldacena13} that this ER=EPR proposal provides a way
to explain the disturbance which occurs in the original firewall paradox
argument\cite{AMPS} when the Hawking radiation is measured without
implying the existence of a firewall.

As a putative counterexample to the firewall paradox, a scenario closely
mimicing the manifestly smooth eternal AdS black hole \cite{Maldacena}
was considered \cite{Maldacena13}, consisting of a pair of `maximally
entangled' black holes connected by an ER bridge. Provided these black
holes evaporate, it is straightforward to extend Theorem 1 to this
scenario: Initially,  there is {\it no\/} firewall, however, by the PT
the paradox is reinstated. No measurements are needed and with no
disturbance, ER=EPR is left without a role. (In fact, the `entanglement'
in this example is local vacuum entanglement across nearby horizons and
vanishes in $O(\text{Planck time})$; see the Appendices).

Notwithstanding this, Ref.~\onlinecite{Susskind14} may be interpreted as
claiming that with the advent of the firewall argument there is at least
something important missing in the conventional AdS/CFT description of
black holes. Indeed, an enduring frustration with the AdS/CFT
correspondence has been that it gives us no inkling of why the `nice
time slice' argument supporting local physics (even prior the PT) is
apparently wrong. As such, it seems wise to guard against blind
acceptance of intuitions about non-local physics coming from holographic
approaches \cite{Maldacena,Hooft,Susskind95}. It is therefore worthwhile
to re\"evaluate the role of non-locality during black hole evaporation.
We do this here with Theorem 2. We shall see that any potential clash
between unitarity and locality actually requires a third ingredient left
out from holographic considerations.

\noindent {\bf Theorem 2}: A contradiction exists between: 2.a)
completely unitarily evaporating black holes, 2.b) large black holes are
described by local physics, and 2.c) externally, a large black hole
should resemble its classical theoretical counterpart (aside from its
slow evaporation).

\noindent {\bf Proof}: Note that assumption 2.a is identical to 
assumption 1.a of theorem 1. 

{\it Local physics} 2.b: In addition to assuming the complete unitary
evaporation of a black hole (2.a), we shall suppose that whatever
process generates the radiation it is constrained to be local for large
black holes. In particular, we shall focus on the fact that local
physics forbids communication across light cones \cite{Strominger}, so
that there can be {\it no\/} communication from within a large black
hole's event horizon to the exterior.

\begin{figure}[ht]
\vskip -0.15truein
\includegraphics[width=52mm]{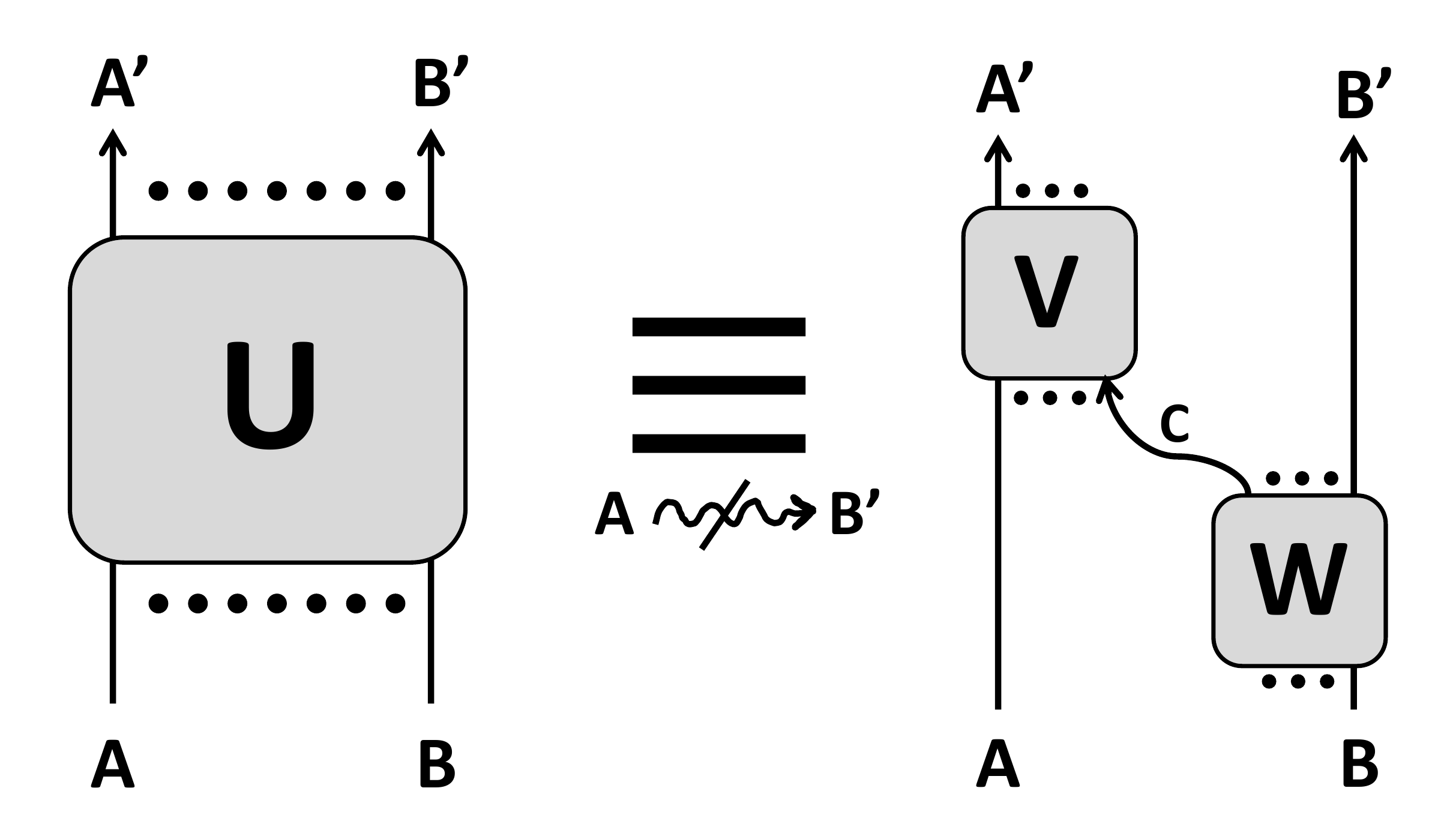}
\vskip -0.18truein
\caption{
Quantum circuit for the no-communication decomposition theorem: Any
unitary process $U$ (left-hand circuit) which maps subsystems $A$ and
$B$ into $A'$ and $B'$ but which does not allow communication from $A$
to $B'$, $A\mathrlap{\kern 0.6em\not}\rightsquigarrow B'$, can be
decomposed into a pair of unitary subprocesses $V$ and $W$ (right-hand
circuit) connected by a `reverse communication' channel $C$.  (The dots
denote ancillary degrees-of-freedom.)
  \label{nocomm}}
\vskip -0.15truein
\end{figure}

In order to make use of this non-communication property we recall the
no-communication decomposition theorem \cite{Werner} (see
Fig.~\ref{nocomm}) which tells us that any unitary process $U$ which
does not allow communication from a set of inputs to a set
of outputs may be decomposed into a pair of unitary subprocesses
$V$ and $W$ with at most some reverse communication within a
subsystem $C$.

\begin{figure}[ht]
\vskip -0.15truein
\includegraphics[width=35mm]{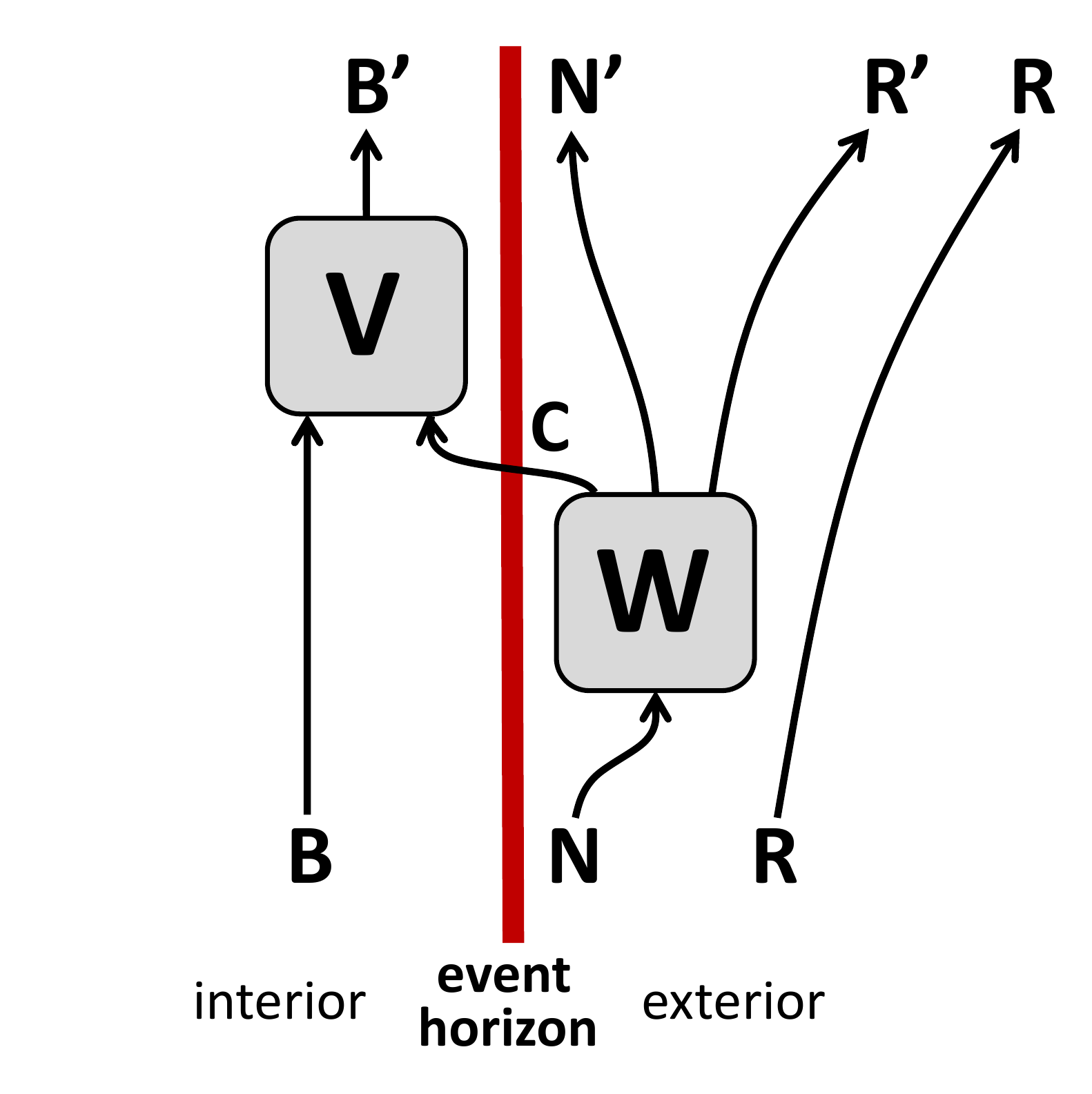}
\vskip -0.18truein
  \caption{
Schematic unitary generation of radiation from a black hole whose
horizon obeys local physics, forbidding communication from interior to
exterior (assumptions 2.a and 2.b, see text). Here $C$ denotes possible
`reverse communication'.
  \label{nocomm_rad}}
\vskip -0.15truein
\end{figure}

This theorem requires that the inputs and outputs form distinct
components of a tensor product decomposition of the overall Hilbert
space; a requirement which is automatic for finite dimensions.  For any
local quantum field theory we may rely on the fact that operators with
support only outside each others light cones must commute. Thus,
locality dictates the existence of the required tensor product structure
across a black hole's horizon \cite{Hawking76,Braunstein13}.  We may
therefore apply the circuit equivalence in Fig.~\ref{nocomm} to the
black hole evaporation of Fig.~\ref{generic_Hawking_process} to give the
structure of an arbitrary unitary black hole evaporation process
consistent with local physics (see Fig.~\ref{nocomm_rad}).

Note that Fig.~\ref{nocomm_rad} is not to be interpreted as a spacetime
diagram. In particular, we do not require that there is any space-like
hypersurface which simultaneously cuts through the subsystems there
displayed. For example, we do not require that subsystem $C$ all arrives
in one block for unitary processing inside the black hole.  From this
perspective, a quantum circuit is a powerful construct.

Strong subadditivity in Fig.~\ref{nocomm_rad} gives
$S(C,N',R'\!:\!R)\ge S(R'\!:\!R)$, and using the unitary
invariance of entropy we have $S(C,N',R'\!:\!R)=S(N\!:\!R)$ leading to
\begin{equation}
S(N:R)\ge S(R':R).
\label{paradox2}
\end{equation}
Note that this inequality involves only correlations between external
degrees-of-freedom and hence relates quantities which are, in principle,
directly observable and reportable. Combining this with the assumption
of complete evaporation, Eq.~(\ref{RpR}), we find
\begin{equation}
{\textstyle \frac{1}{2}}
S(N:R)\ge 
\left\{
\begin{array}{lr}
S_{\text{BH}}-\log_e |R'|, & \text{pre-PT},\\
\log_e |R'|, & \text{post-PT}.
\end{array}
\right.
\label{RadENT3}
\end{equation}

{\it Non-exotic atmosphere} 2.c: We shall take assumption 2.c to be
equivalent to 1.b(i), that the exterior should not consist of exotic
matter. The holographic entropy bound \cite{Bousso99} shows that an
entropy of at most $O({\cal A}_{\text{horizon}}^{1/2})$ can reside
between the causal and stretched horizons, so their distinction has
negligible effect on our analysis. For a partially evaporated black hole
with Bekenstein Hawking entropy $\log_e |R'|$,
to ensure a non-exotic atmosphere, Eq.~(\ref{Correspondence}) requires
$\log_e |R'|\gg S_{\text{therm}}^N \ge\frac{1}{2}S(N\!:\!R)$. Combining
this with Eq.~(\ref{RadENT3}) yields
\begin{equation}
\log_e |R'|\gg\!
{\textstyle \frac{1}{2}} S(N\!:\!R)\ge\!
\left\{\!
\begin{array}{lr}
S_{\text{BH}}-\log_e |R'|, & \!\!\!\!\text{pre-PT},\\
\log_e |R'|, & \!\!\!\!\text{post-PT}.
\end{array}
\right.
\label{RadENT4}
\end{equation}
Except for the very earliest stages of evaporation, this result yields
a contradiction.
{$~$}\hfill \rule{2mm}{2mm}

\vskip 0.03truein

{\it Discussion:} Both theorems apply to the behavior of large black
holes where General Relativistic reasoning is conventionally expected to
hold. Theorem 1, which yields a contradiction from the Page time (PT)
onward, ignores the local structure of a black hole's horizon and
suggests that huge thermodynamic entropies reside within the black hole.
Theorem 2, which yields a contradiction almost immediately once
evaporation has begun, incorporates this local structure and instead
suggests that huge entropies reside external to the event horizon. 

The cheapest resolution, cut along the lines of Occam's razor, would be
to reject assumption 1.a (2.a). However, for unitarity to be preserved,
black hole evaporation cannot stop when some `stable remnant' is reached
\cite{Hooft,Bekenstein94,Giddings95}, nor can the black hole interior
`bud off' as a baby universe \cite{Hooft,Banks84}. Any such loss of
unitarity would infect almost every other quantum mechanical process
\cite{Hooft,Giddings13}. Unfortunately, as already noted, the firewall
paradox has precipitated a crisis for the AdS/CFT correspondence
\cite{Maldacena13}, so this route to ensuring unitary black hole
evaporation lies on uncertain ground, at least until a suitable
dictionary can be found between the CFT and black hole interior states.
Indeed, our analysis seems to imply that such a dictionary must be
dynamic, varying with the black hole's age.

If we accept 1.a (2.a), either theorem leaves us with a striking
dichotomy. Let us start with the consequences of Theorem 1. We must
reject at least one assumption of 1.c or 1.b. To start with, were the
accessible dimensionality within the stretched horizon larger than the
estimate given by the Bekenstein Hawking entropy, we would be able fill
a black hole with more thermodynamic entropy than could be accounted for
by the entropy that would eventually appear as radiation. Thus, if
assumption 1.c failed to hold the theory of black hole thermodynamics,
and possibly thermodynamics itself, could {\em not} survive.

By contrast, assumption 1.b, although usually considered a consequence
of the Equivalence Principle of General Relativity is in fact nothing
more than a boundary condition on the quantum fields at the event
horizon; it is well known that different choices of `vacuum state' lead
to wildly different behaviors for the energy-momentum tensor there.
Splitting 1.b into its components: A failure of 1.b(i), would imply that
the exterior must consist of super-entropic exotic matter (such as an
atmosphere of microscopic black holes), and so would almost certainly
have some observational consequences \cite{Telescope}. Finally, a
failure of 1.b(ii) would imply that by the PT a black hole's interior is
filled with half of a maximally entangled state.

In 2009 it was noted that maximal entanglement between the black hole
interior and the radiation implied an absence of entanglement across the
horizon \cite{Braunstein13}: ``In an arbitrary system where
trans-boundary entanglement has vanished, the quantum field cannot be in
or anywhere near its ground state. Applied to black holes, a loss of
trans-event horizon entanglement implies fields far from the vacuum
state in the vicinity of the event horizon.''

We now extend this reasoning: As there is {\em no} entanglement
between {\em any} pieces of the quantum fields within the black hole, no
place within the interior can look like a low-energy vacuum state
--- like regular spacetime. We might say, in this sense, that from the
PT onward there is {\em no spacetime} within the black hole.

Next, consider the options left by Theorem 2: to reject at least one of
the assumptions 2.c or 2.b. Rejecting 2.c is equivalent to rejecting
1.b(i). Therefore, the only other minimal option (rejecting assumption
2.b) would be to assume that communication from the black hole interior
to exterior across the horizon was possible. In particular, one might
note that a ``tunneling'' mechanism has been long anticipated to provide
a more powerful explanation for black hole radiation
\cite{Manas,Parikh00,Braunstein13}. However, tunneling across the
horizon alone\cite{Parikh00} is insufficient, as it still leaves
unanswered how the degrees-of-freedom from deep within a black hole
manage to (non-locally) reach up to just inside the horizon where they
can participate in such tunneling. Our mechanism (below) supporting the
firewall phenomenon may explain how this can occur.

{\it Simplifying our assumptions}: It turns out that there are several
ways of reducing the assumptions needed to obtain {\em Theorem 1}.
First, we may drop assumption 1.c entirely if we can apply the
holographic entropy bound \cite{Bousso99} within a black hole's horizon.
Recall the converse interpretation of this bound which states that the
minimal area encompassing a given thermodynamic entropy is four times
that entropy (in Planck units). The covariant form of this bound
\cite{Bousso99} should apply anywhere (including within a black hole's
horizon). Hence, let ${\cal A}_{\text{min}}$ be the minimal area
encompassing a thermodynamic entropy of $\frac{1}{2}S(B\!:\!R)$. By
assumption 1.b, this area must be well within the (surface area of the)
horizon, implying that $\log_e|R'| \gg {\textstyle \frac{1}{4}}{\cal
A}_{\text{min}} \ge {\textstyle \frac{1}{2}}S(B\!:\!R)$. This result
directly contradicts Eqs.~(\ref{RpR}) and~(\ref{BR}). Interestingly, if
we conservatively assume that this area is centrally located, so that
assumption 1.b holds for as long as possible, then an observer freely
falling from infinity will hit the firewall almost immediately upon
crossing the horizon.

Finally, let us focus on the onset of the paradox: Consider a
black hole, created from matter in a pure quantum state, which unitarily
evaporates at least until the Page time. We shall only suppose that the
net (von Neumann) entropy in the radiation is equal to the Bekenstein
Hawking entropy of the black hole at that stage. We then trivially have
$\frac{1}{2}S(B,N\!:\!R)=\frac{1}{2}S_{\text{BH}}$. Therefore, if this
black hole is to be free of an exotic atmosphere at the Page time, then
virtually all this thermodynamic entropy must reside inside, out of
view. Consistency with black hole thermodynamics (or the holographic
entropy bound) then dictates that there is no room left within the black
hole for a visitor to keep her cool!

{\it Speculation}: We end with a purely speculative description for how
black holes might evolve during evaporation and consequently outline a
{\it mechanism\/} supporting the firewall phenomenon, as suggested by
our work. During the initial stages of evaporation, prior to the Page
time (PT), entanglement grows between the interior and distant
radiation. Now entanglement cannot be compressed into fewer qubits than
given by its entropy (a principle which in no way assumes that the
associated matter has been compressed to Planck densities). Therefore,
accepting an effectively finite size Hilbert space to the black hole
interior (or applying the holographic entropy bound there), the interior
slowly fills up with {\it incompressible\/} entanglement. This
incompressible `stuff' would grow outward from what would classically be
the black hole singularity, while simultaneously, the black hole's
horizon is shrinking as radiation is emitted.

We conjecture therefore that there is some well-defined internal {\it
entanglement surface} that contains the entanglement growing outward. At
the PT the entanglement surface and black hole horizon meet. At that
stage the horizon may survive, or may be replaced by the entanglement
surface. In the former case, evaporation would continue by something
very much like quantum tunneling \cite{Parikh00,Braunstein13,Manas} from
degrees-of-freedom on the entanglement surface just inside the horizon.
In the latter case, evaporation may continue via direct ejection from
the {\it entanglar\/} (entangled star) though its detailed spectrum
(e.g., its neutrino flux) and lifetime would almost certainly differ
from a true black hole with otherwise identical mass, charge and angular
momentum. Na\"ively, an entanglar (of even a modest size) would take far
longer than the age of the universe to evolve from a black hole, so none
can be expected to currently exist. Conversely, the unambiguous
observation of such an entanglar would yield {\it prima facie\/}
evidence for an object that far predates the Big Bang.

\section*{Appendices}

\subsection*{The ER=EPR `counterexample'}

{\em Classical 3-Manifold structure}: We start by considering the prime
counterexample considered in the ER=EPR proposal \cite{Maldacena13}.
This consists of a pair of black holes connected by an Einstein Rosen
bridge. (Physically, this might correspond to what is produced by a
pair-creation event.)

\begin{figure}[th!]
\includegraphics[width=55mm]{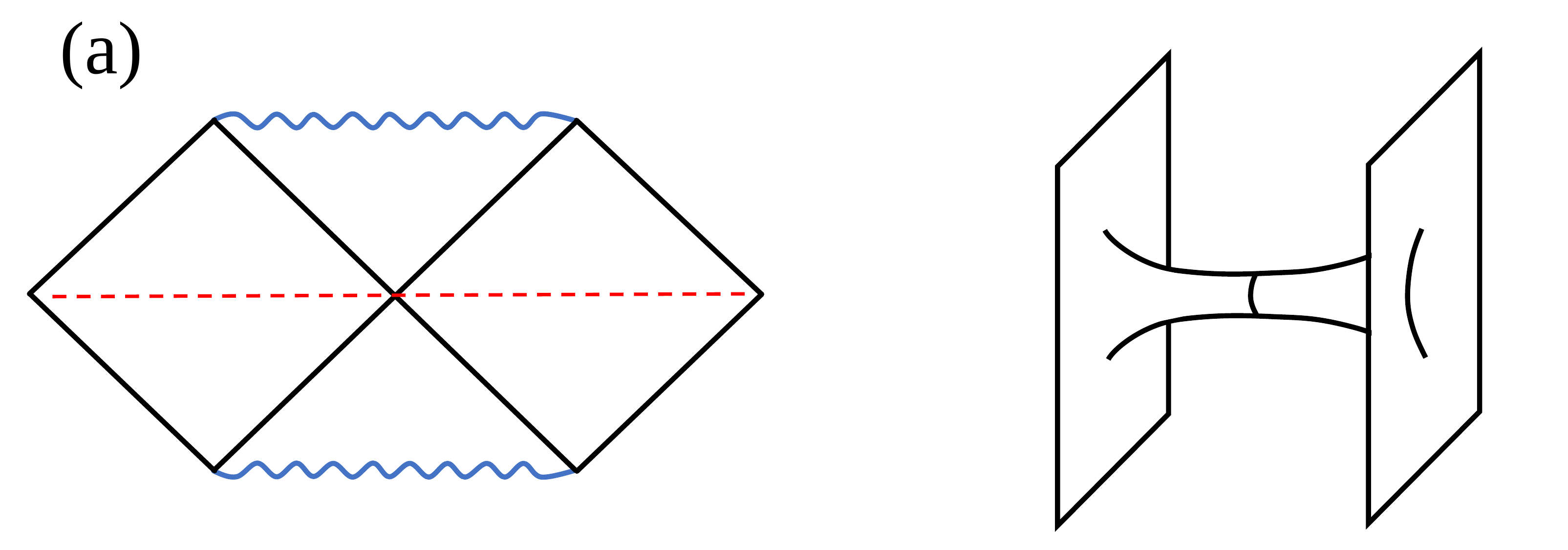}
\vskip -0.1in
\includegraphics[width=55mm]{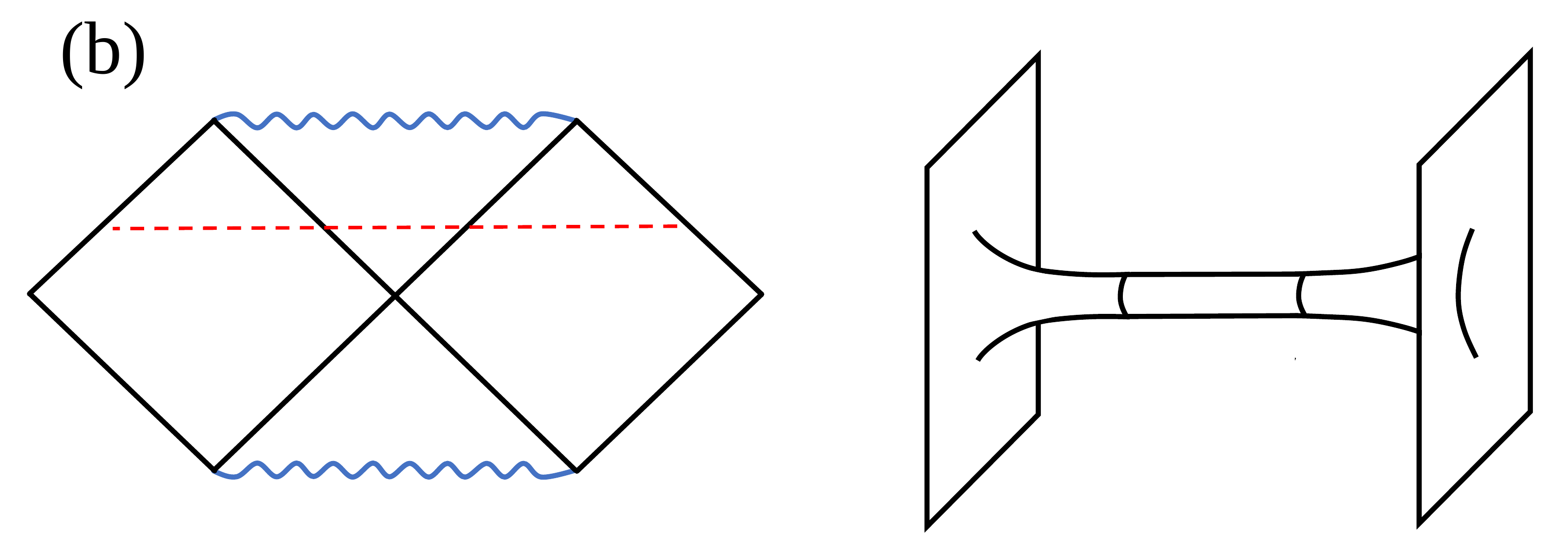}
\vskip -0.1in
  \caption{
Structure of a pair of black holes connected by an Einstein Rosen (ER)
bridge. The dashed red lines on the Penrose diagrams (left) denote
the space-like hypersurfaces used to construct the embedding diagrams
(right). Case: (a) when the pair are initially formed (say by a
pair-creation event) the horizons coincide; and (b) at a
later time; the horizons are separating, through the ER bridge, at the
speed of light.
\label{Fig4}
}
\end{figure}

If we ignore evaporation, the black hole exteriors are static and
eternal. Their joint Penrose diagram is shown in Fig.~\ref{Fig4} as the
left and right black diamond shapes of either of the left-hand figures.
Each dashed red line denotes some specific space-like hypersurface (a
specific time slice). In Fig.~\ref{Fig4}(a), this time slice is chosen
when the two exteriors touch at the center of the left-hand figure. The
black $45\degree$ diagonal lines passing through the center of this
figure represent the horizons of the two black holes respectively. With
regard to the scenario where these black holes are created by some
pair-creation process, this ``$t=0$'' time slice would correspond to
their initial creation event, and the Penrose diagram loses any meaning
for earlier times. The right-hand diagram is the spatial embedding
diagram corresponding to this initial time slice. Far from either black
hole, externally, the embedding diagram looks flat. As one approaches
either black hole from the outside, one approaches a horn-like structure
on the embedding diagram. The horn structure terminates at the horizon,
denoted as a vertical black ring that encircles the horn. For the $t=0$
hypersurface the horizons of the two black holes coincide.

Fig.~\ref{Fig4}(b) shows the same black hole pair, but a later
hypersurface is chosen (left-hand figure). The corresponding embedding
diagram (right-hand figure) shows that the two horizons have separated
and are connected internally by a bridge --- the so-called
Einstein-Rosen (ER) bridge. The proper length of this bridge grows very
rapidly (at roughly the speed of light) so there is no possibility of
passing from the exterior of one black hole to the exterior of the
other. The pair forms an example of non-traversable wormhole.

\begin{figure}[b!]
\includegraphics[width=60mm]{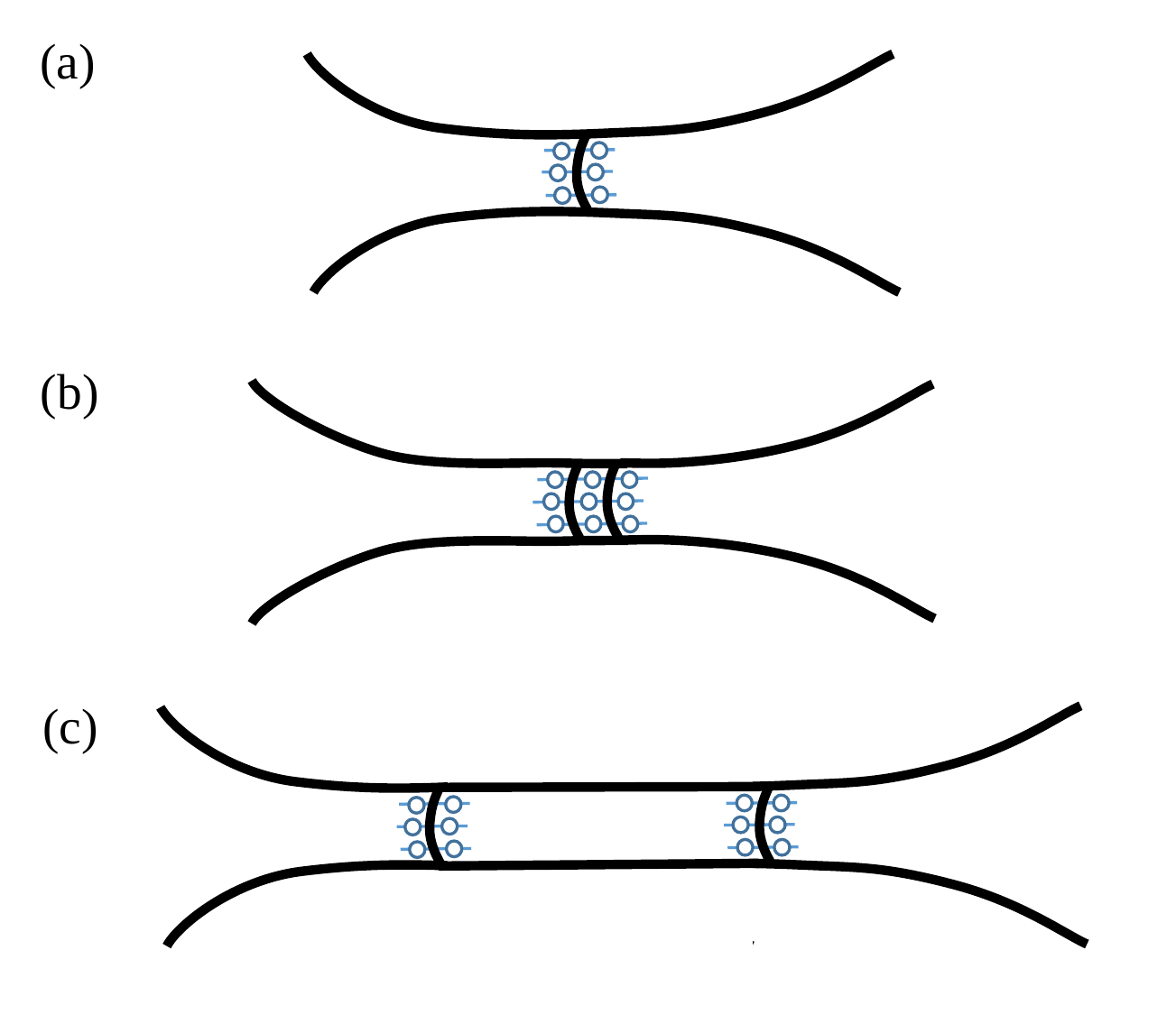}
\vskip -0.1in
  \caption{
Schematic description of entanglement across the horizons for our black
hole pairs. Entanglement is sketched on the embedding diagrams of the
3-manifolds based on a field theoretic formulation on a lattice. Each cell
(small blue circle) denotes an individual lattice site. For clarity,
only those cells neighboring the horizons are shown. Entanglement across
the horizons is shown by (light blue) lines connecting cells at three
epochs: (a) the initial hypersurface when the horizons coincide; (b)
when the horizons are separated by $O({\text{Plank length}})$; and (c)
on a later hypersurface.
\label{Fig5}
}
\end{figure}

{\em Quantum fields and entanglement}: Continuing to ignore evaporation,
we can consider quantum fields propagating on the time-evolving family
of spatial 3-manifolds corresponding to the family of embedding diagrams
for different hypersurfaces (time slices). Provided we stay away from
the singularity (wavy blue line at the top of the Penrose diagrams) all
these 3-manifolds are smooth and locally flat. The lowest energy states
(vacuum) of these quantum fields will then not be too different from the
structure of vacuum in flat spacetime. In particular, there will be
entanglement across all boundaries. This may be formalized, for example
on a quantum field theoretic setting on a lattice \cite{BDS13,Braunstein13},
with the typical lattice spacing determining the UV cutoff as Planckian.
It is sufficient for our purposes here to consider schematic pictures of
the entanglement on such a lattice representation of these 3-manifolds.

Fig.~\ref{Fig5} shows such a schematic representation of entanglement on
these 3-manifolds. The lattice sites are represented as small blue
circles. For clarity, only those circles neighboring horizons are shown.
Entanglement across the horizons is shown as pale blue lines connecting
lattice sites. As can be seen in Fig.~\ref{Fig5}(a) the initial $t=0$
hypersurface does indeed show entanglement across the joint horizon. The
exteriors of the two black holes are indeed entangled on this time
slice. 

In Fig.~\ref{Fig5}(b), we see the entanglement for a hypersurface at
$t=O({\text{Planck time}})$. Once the horizons have separated by even of
order one lattice site, presumed to be separated by $O({\text{Planck
length}})$, the entanglement between the black hole exteriors includes a
set of intermediary lattice sites. When these intermediary sites are
traced out the original entanglement will almost have vanished. As the
separation between the horizons increases the entanglement between the
black holes is exponentially suppressed, effectively vanishing. Thus,
on a later hypersurface, such as Fig.~\ref{Fig5}(c), there will be no
entanglement between the black holes.

We may therefore conclude, (i) that on the initial $t=0$ hypersurface
there is indeed entanglement of the external degrees-of-freedom for the
black hole pair. However, (ii) this vanishes within $O({\text{Planck
time}})$. Further, (iii) being local entanglement across a horizon this
has no observable consequences. For such a short-lived phenomenon one
might question whether this entanglement is perhaps better thought of as
a mathematical artifact.

Before we proceed to seeing how theorem 1 applies to this counterexample
we might consider other ways of creating maximally entangled pairs of
black holes. Indeed, Ref.~\onlinecite{Maldacena13} suggests other
mechanisms by which the {\em internal} degrees-of-freedom of a pair of
black holes may be maximally entangled. For example, by waiting for one
black hole to radiate until its Page time and then collapse the
resulting radiation into a second black hole. However, all of these
alternative suggested mechanisms have entanglement of a completely
different character than the counterexample studied above. The
entanglement is not ephemeral and it is between the internal instead of
exterior degrees-of-freedom. Thus, there is no connection between the
smoothness or otherwise of the quantum fields for these mechanisms and
the counterexample above.

{\em Theorem 1 for the ER=EPR `counterexample'}: Finally, we shall
consider evaporation in the scenario of pair-created black holes studied
above. To apply theorem 1, all be need to do reinterpret $B$, $N$, $R$,
etc as the Hilbert spaces of the joint interior, the combined
neighborhoods and combined radiation systems for the black hole pair.
Assumption 1.c needs to be modified to ``the joint black interior Hilbert
space dimensionality may be well approximated as the exponential of the
combined Bekenstein Hawking entropy of the black hole pair''. As noted
in Fig.~1 of the manuscript, the quantum state of $(B,N,R)$ is
arbitrary. All the equations used to derive the contradiction for
theorem 1 remain unchanged. We find the same paradox as before, with its
onset at the Page time, when the joint surface area of the black hole
pair has dropped to one-half its initial value. 

The proposed counterexample to the paradox thus fails.

\subsection*{Post-firewall paradoxes with negligible entropies}

In this section we repeat the key elements of both theorems in the
manuscript with the following modifications: (a) We explicitly include
the entropy in the atmosphere, bounding its size rather than merely
considering it to be negligible; (b) We only follow the black hole
evaporation to the point where the black hole is still much larger than
Planck scale. To illustrate that neither of these changes affect the
results of the manuscript we focus solely on the behavior at the
Page time.

{\em Theorem 1}:

Consider now the scenario where we follow a black hole to a relatively
late stage of its complete evaporation.  In particular, when its area
has shrunk to some small fraction of its original size, but is still
much larger than the Planck scale so that the physics of Planck scale 
black holes plays no part. We denote all pre-Page time radiation as $R$
and the post-Page time radiation as $R'$ (produced up until the black
hole has reached a specified fraction, say roughly $\varepsilon/2$,
of its original area). It follows therefore from the generic behavior 
of entropy during evaporation \cite{Braunstein13} that
\begin{equation}
S(R':R)=(1-\varepsilon)S_{\text{BH}}, \qquad \varepsilon\ll 1.
\label{RadENT}
\end{equation}
Combining this with Eq.~(2) of the manuscript we find
\begin{equation}
S(B,N:R)\ge(1-\varepsilon)S_{\text{BH}}, \qquad \varepsilon \ll 1.
\label{RadENT2}
\end{equation}

Equation~(\ref{RadENT2}) tells us that the radiation is almost perfectly
maximally entangled with a subspace of the joint system $(B,N)$ and as
$R$ quickly becomes remotely separated we may conclude that
$\frac{1}{2}(1-\varepsilon)S_{\text{BH}}$ represents a lower bound to
the thermodynamic entropy of this joint system.

{\it Free-fall equanimity}: Consider now a freely-falling observer who
is believed to see nothing special until they pass well within a large
black hole's horizon (assumption 1.b). For black holes formed by matter
in a pure quantum state, the (global) state of $(B,N,R)$ may also be
treated as pure implying $S(B,N\!:\!R)= S(B\!:\!R)+S(N\!:\!R)$. This in
turn, allows assumption 1.b to be decomposed into external and internal
constraints.

Externally, we assume that our infalling observer is not passing through
an atmosphere of exotic matter prior to reaching the horizon. Therefore
from Eq.~(1) of the manuscript, we have $\frac{1}{2}S(N\!:\!R) \ll
\frac{1}{2} S_{\text{BH}}$ for a black hole at the Page time.

Internally, this implies that Eq.~(\ref{RadENT2}) reduces to
\begin{equation}
S(B:R)\gtrsim S_{\text{BH}}.
\label{reduced}
\end{equation}
Now, a trivial bound to the quantum mutual information is
that $2\log_e({\text{dim}}(B))\ge S(B\!:\!R)$. If this bound were
saturated, the huge thermodynamic entropy inside $B$ would imply that an
infalling observer would {\it immediately\/} encounter an incredibly
mixed state (e.g., a near uniform mixture of roughly $10^{10^{77}}$
orthogonal quantum states for an initially stellar mass black hole) with
correspondingly huge energies as soon as they passed the horizon. They
would immediately encounter an `energetic curtain' \cite{Braunstein13}
or firewall \cite{AMPS} upon entering the black hole. To guarantee
assumption 1.b holds, the above dimensional bound must be far from
saturation, i.e., at the Page time
\begin{equation}
\log_e( {\text{dim}}(B))\gg \frac{1}{2}\,S_{\text{BH}}.
\label{combined1ab}
\end{equation}

{\it Finite interior Hilbert space}: We may now derive a
contradiction along the lines of the original firewall paradox.
Assumption 1.c holds that the black hole interior has a Hilbert space
dimensionality that is well approximated by the exponential of the
Bekenstein-Hawking entropy. Thus, at the Page time, when a black hole's
surface area has shrunk to one-half of its original value we would have
\begin{equation}
\log_e( {\text{dim}}(B))\simeq
\frac{1}{2}\,S_{\text{BH}},
\label{BHentropy}
\end{equation}
which directly contradicts Eq.~(\ref{combined1ab}).
{$~$}\hfill \rule{2mm}{2mm}
\vskip 0.05truein

{\em Theorem 2}:

Note that Eq.~(6) of the manuscript involves only correlations between
external degrees-of-freedom and hence relates quantities which are, in
principle, directly observable and reportable. Combining this with the
assumption of complete evaporation, Eq.~(\ref{RadENT}), we easily find
\begin{equation}
S(N:R)\ge (1-\varepsilon) S_{\text BH}, \qquad \varepsilon\ll 1.
\label{RadENT3}
\end{equation}

Locality (assumption 2.b) has allowed us to eliminate $B$ from
Eq.~(\ref{RadENT2}), which in turn allows us to do without any specific
bound to the size of the interior Hilbert space. More surprisingly,
locality implies a very different picture: one where huge thermodynamic
entropies must reside outside the black hole instead of inside it. 

At first sight, this appears reminiscent of arguments based on
time-reversing Hawking radiation. Ordinary Hawking radiation evolves out
of vacuum modes, but any (information bearing) deviations were argued to
have started out as high-energy excitations near the horizon
\cite{Giddings94}. By contrast, the huge entropies in
Eq.~(\ref{RadENT3}) are associated with degrees-of-freedom that are
maximally entangled with the outgoing radiation and therefore correspond
to an effect of the ``infalling partners'' to the radiation. Thus,
Eq.~(\ref{RadENT3}) represents a distinct (and much stronger) phenomenon 
imposed by locality.

{\it Non-exotic atmosphere}: Assumption 1.c is weaker than 1.b,
only requiring that externally, black holes should resemble their
classical counterparts (aside from their slow evaporation). In turn, we
apply this in a weak manner to only suppose that the black hole
does not contain an atmosphere of super-entropic exotic matter.
From Eq.~(1) of the manuscript
\begin{equation}
S(N:R) \le \eta\, S_{\text{BH}}, \qquad \text{with $\eta\ll 1$},
\label{EPeqn}
\end{equation}
and combining Eqs.~(\ref{RadENT3}) and~(\ref{EPeqn}) yields the
contradiction
\begin{equation}
1\le \varepsilon+\eta\ll 1,
\end{equation}
whatever the details of the radiation process.
{$~$}\hfill \rule{2mm}{2mm}

\section*{Arbitrary infallen matter}

In this section we generalize our results to show that they
apply even when the matter that collapsed to form the black hole
is not pure. We start with a more general review of generic
black hole radiation necessary to analyse such scenarios.

\subsection*{Generics of black hole radiation}

\vskip -0.1truein

In the manuscript we considered a black hole with (initial)
thermodynamic entropy $S_{\text{BH}}$ which can completely evaporate
into a net pure state of radiation. As discussed, the generic
evaporative dynamics of such a black hole may be captured by Levy's
lemma for the random sampling of subsystems from an initially pure state
consisting of $S_{\text{BH}}$ qunats \cite{Braunstein13}. This either
assumes the infallen matter is pure (as in the manuscript) or ignores it
entirely. Throughout, we set Boltzmann's constant to unity and work with
natural logarithms leading to the measure of qunats (i.e., $\log_e 2$
times the number of qubits \cite{bit}).

In order to extend our analysis to include infallen matter carrying some
(von Neumann) entropy $S_{\text{matter}}$, we need only take the
initially pure state used above and replace it with a bipartite pure
state consisting of two subsystems: $S_{\text{BH}}$ qunats to
represent the degrees-of-freedom that evaporate away as radiation; and a
reference subsystem. Without loss of generality, the matter's entropy
may be treated as entanglement between these two subsystems, however,
here we shall simplify our analysis by assuming uniform entanglement
between the black hole subsystem and $S_{\text{matter}}$ reference
qunats. The generic properties of the radiation may then again be
studied by random sampling the former subsystem to simulate the
production of radiation \cite{Braunstein13}.

The behavior is generic and for our purposes may be summarized in terms
of the radiation's von Neumann entropy, $S(R)$, as a function of the
number of qunats in this radiation subsystem. One finds
\cite{Braunstein13} that $S(R)$ initially increases by one qunat for
every extra qunat in $R$, until it contains
$\frac{1}{2}(S_{\text{BH}}+S_{\text{matter}})$ qunats. From that stage
on it decreases by one qunat for every extra qunat in $R$ until it drops
to $S_{\text{matter}}$ when $R$ contains $S_{\text{BH}}$ qunats and the
black hole has completely evaporated.

Because the von Neumann entropy of a randomly selected subsystem only
depends on the size of that subsystem, the same behavior is found whether
$R$ above represents the early or late epoch radiation with respect
to any arbitrary split. Further, in the simplest case where we choose
the joint radiation $(R,R')$ to correspond to the net radiation from
a completely evaporated black hole we may immediately write down the
generic behavior for the quantum mutual information $S(R':R)$.

In particular, $S(R':R)$ starts from zero when $R$ consists of zero
qunats.  From then on, it increases by two qunats for every extra qunat
in $R$ until $S(R':R)$ reaches $S_{\text{BH}}-S_{\text{matter}}$ when
$R$ contains $\frac{1}{2}(S_{\text{BH}}-S_{\text{matter}})$ qunats. From
that stage on until $R$ contains
$\frac{1}{2}(S_{\text{BH}}+S_{\text{matter}})$ qunats $S(R':R)$ remains
constant, after which $S(R':R)$ decreases by two qunats for every extra
qunat in $R$ until it drops to zero once the $R$ contains the full
$S_{\text{BH}}$ qunats of the completely evaporated black hole
\cite{Braunstein13}. Interestingly, it is during the region where
$S(R':R)$ is constant that the information about the infallen matter
becomes encoded into $R$ for the first time \cite{Braunstein13}.
Finally, setting $S_{\text{matter}}$ to zero gives the `standard'
behavior for $S(R)$ and $S(R'\!:\!R)$ upon which the results in the
manuscript are derived.

From the above, we are motivated to generalize the Page time: we define
{\it any\/} time where $S(R'\!:\!R)$ is maximal a (generalized) Page
time; the earliest such time the `initial Page time'; and the latest the
`final Page time'. Prior to the initial Page time, the quantum
information about the initial infallen matter is encoded entirely within
the black hole interior \cite{Braunstein13}. After the final Page time
this information is encoded entirely within the radiation
\cite{Braunstein13}.

\vskip -0.4truein

\subsection*{Including infallen matter}

\vskip -0.1truein

Let us start with a consideration of how the reasoning in Theorem 2
becomes modified by the presence of infallen matter carrying entropy.

\begin{figure}[h!]
\vskip -0.1in
\includegraphics[width=38mm]{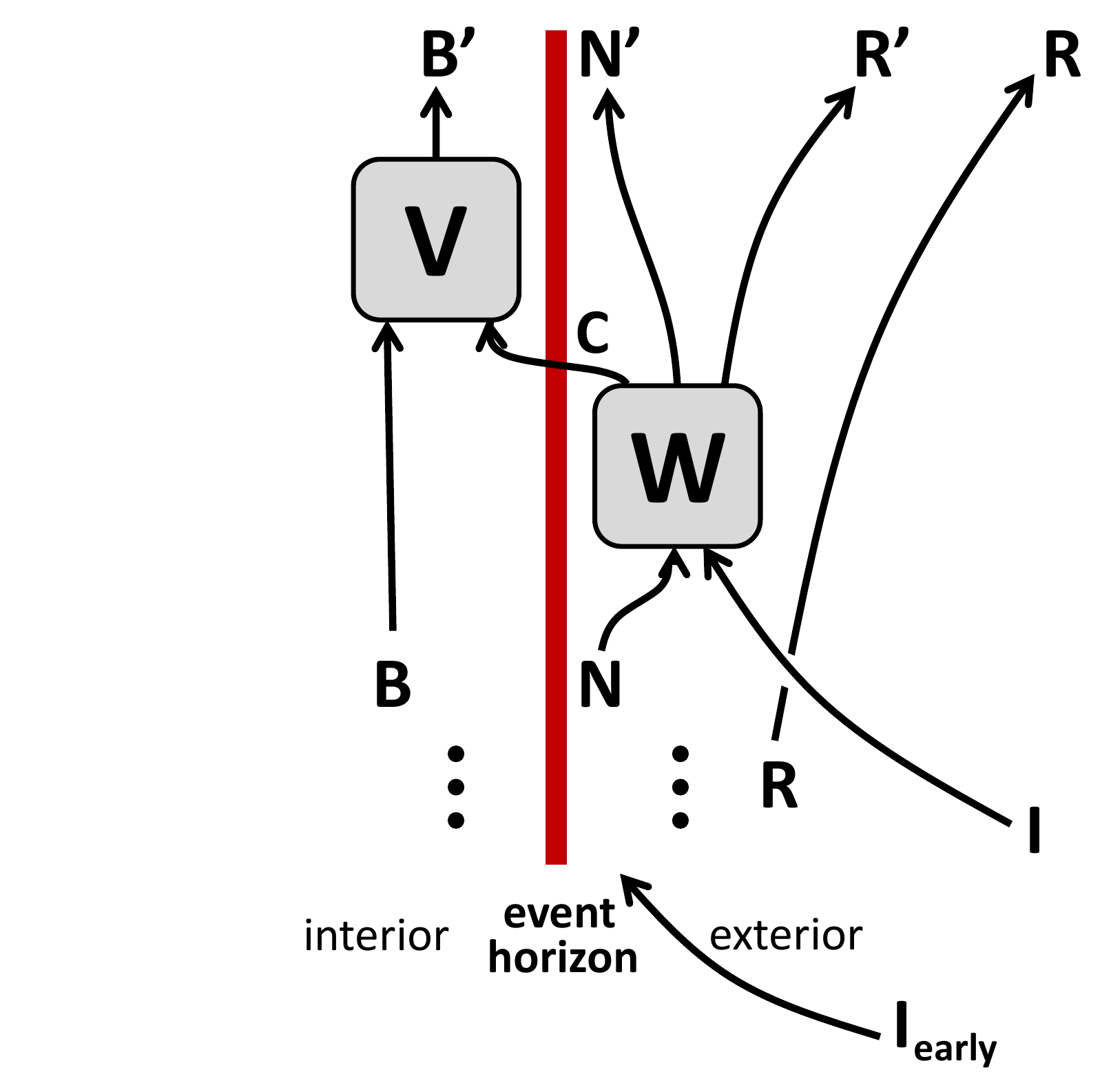}
\vskip -0.15in
  \caption{
Quantum circuit diagram for evaporation of a quantum black hole with
causal horizon and infallen matter.  Subsystem $I$ denotes infallen
matter that falls into the region surrounding the black hole to
participate in late epoch radiation generation.  (This does not exclude
the possibility that the matter falls directly into the black hole.)
Subsystem $I_{\text{early}}$ denotes matter infalling at earlier times
or even that collapses to form the original black hole.
\label{Fig6}
}
\end{figure}

\noindent {\bf Theorem 2 generalized}: In the main body of the
manuscript we did not explicitly include entropy associated with
infallen matter.  Fig.~\ref{Fig6} shows the most general scenario.
Subsystem $I$ denotes the matter that falls into the region surrounding
the black hole where radiation is produced. Thus, we suppose that late
epoch radiation can in principle come from the joint subsystem $(N,I)$.
In this figure we also include subsystem $I_{\text{early}}$ denoting
matter that has fallen into the region surrounding the black hole at an
earlier epoch or indeed matter that may have collapsed to form the black
hole in the first place.

As in the manuscript we apply strong subadditivity:
\begin{eqnarray}
S(R':R) &\le& S(C,N',R':R) \nonumber \\
&=& S(N,I:R) = S(N:R).
\label{NewMain1}
\end{eqnarray}
Here, we used the fact that joint subsystems $(C,N',R')$ and $(N,I)$ are
unitarily related. Finally, the most natural assumption is that the
infallen matter $I$ is independent of the quantum state of the black
hole, $(B,N)$, or its early epoch radiation $R$. The original inequality
of Eq.~(10) from the manuscript is thus found to still hold in the presence
of infallen matter.

From the summary above of generic radiation production including
infallen matter we have enough to generalize Theorem 2. As in the
manuscript, we take $R$ to be all the early epoch radiation until the
Page time (for this theorem we may take any generalized Page time), and
we let $R'$ denote all the radiation generated from the Page time
onward until the black hole has shrunk to a size much smaller than the
original black hole (say roughly $\varepsilon/2$ of its original area),
but still much larger than the Planck scale. In this case, instead of
Eq.~(\ref{RadENT}), we have
\begin{equation}
S(R':R)= (1-\varepsilon) \, S_{\text{BH}}-S_{\text{matter}},
\qquad \varepsilon\ll 1,
\label{RadCorrMatter}
\end{equation}
where $S_{\text{matter}}\equiv S(I_{\text{early}},I)$ is the net entropy
in all the infallen matter. Combining this with Eqs.~(\ref{EPeqn})
and~(\ref{NewMain1}) gives
\begin{equation}
1 -\frac{S_{\text{matter}}}{S_{\text{BH}}} \le \epsilon + \eta \ll 1.
\end{equation}
Once again we obtain a contradiction except in the extreme case of a
black hole whose net infallen matter contains virtually as much entropy
as the entire black hole's original entropy
$S_{\text{BH}}$.

\vskip 0.1truein

\noindent
{\bf Theorem 1 generalized}:
It is simple enough to repeat the above reasoning for Theorem 1, where
we no longer make use of locality. In this case, we may still use
Fig.~\ref{Fig6} provided we ignore the no-communication
decomposition structure. In particular, strong subadditivity yields
\begin{eqnarray}
S(R':R) &\le& S(B',N',R':R) \nonumber \\
&=& S(B,N,I:R) = S(B,N:R),
\label{NewMain2}
\end{eqnarray}
which is identical to Eq.~(2) of the manuscript. Here, we use the fact
that joint subsystems $(B',N',R')$ and $(B,N,I)$ are unitarily related.
Again, the most natural assumption is that the infallen matter $I$ is
independent of the quantum state of the black hole, $(B,N)$, or its
early epoch radiation $R$.

Applying Eq.~(\ref{RadCorrMatter}) to any Page time then tells us that
for a unitarily and completely evaporating black hole
\begin{equation}
S(B,N:R) \ge (1-\varepsilon) \, S_{\text{BH}}-S_{\text{matter}},
\qquad \varepsilon \ll 1.
\end{equation}

To simplify our argument, we shall suppose that the infallen matter
($I_{\text{early}}, I)$ has actually entered the black hole. In that
case, for any times prior to the initial Page time, the infallen
matter's external reference qunats are maximally entangled with some
subsystem of the black hole interior \cite{Braunstein13}. We shall label
the orthogonal complement of this subsystem within $B$ as $B^{\perp}$.
It is clear that: i) $(B^{\perp},N,R)$ can be treated as a pure quantum
state; and ii) $S(B,N\!:\!R)=S(B^{\perp},N\!:\!R)$. So that
\begin{equation}
S(B^{\perp}:R)+S(N:R) \ge (1-\varepsilon) \,
S_{\text{BH}}-S_{\text{matter}}, \qquad \varepsilon \ll 1.
\label{xxx}
\end{equation}
To ensure that our infalling observer is not passing through an 
atmosphere of exotic matter before they reach the horizon, Eq.~(1)
from the manuscript for a large black hole implies that Eq.~(\ref{xxx})
reduces to
\begin{equation}
S(B^{\perp}:R) \gtrsim
S_{\text{BH}}-S_{\text{matter}}.
\end{equation}
Since $\log_e ({\text{dim}}(B^{\perp}))=\log_e ({\text{dim}}(B))-
S_{\text{matter}}$ by construction, we find the trivial bound
\begin{equation}
\log_e({\text{dim}}(B)) \gtrsim \frac{1}{2}(S_{\text{BH}}+S_{\text{matter}}).
\end{equation}
If this bound were saturated, then the huge thermodynamic entropy in
$B$ would imply that an infalling observer would immediately encounter an
incredibly mixed state with correspondingly huge energies as soon
as they passed the horizon. They would immediately encounter
an `energetic curtain' or firewall upon entering the black hole.
To ensure, therefore that assumption 1.b holds, the above bound must
be far from saturation, i.e.,
\begin{equation}
\log_e({\text{dim}}(B)) \gg \frac{1}{2}(S_{\text{BH}}+S_{\text{matter}}),
\label{gg}
\end{equation}
where $B$ is the black hole at the {\it initial\/} Page time.

However, assumption 1.c would require that the left- and
right-hand-sides of Eq.~(\ref{gg}) should be nearly equal. As with the
generalization of Theorem 2, we again obtain a contradiction, in this
case, however, apparently independent of the amount infallen matter.


\begin{thebibliography}{99}
\bibitem{fn3}
This subspace of $X$ must have Schmidt coefficients consistent with the
entanglement and hence carrying entropy $\frac{1}{2}S(X\!:\!Y)$. By
mimicing this subspace with a locally created probabilistic state given
by $\text{tr}_Y(\rho_{XY})$ every local property, including
thermodynamic, can be seen to be identical in the original and mimiced
states.
\bibitem{fnPir}
Note, Levy's lemma shows that mechanisms with behavior different from
the generic will be exponentially unstable to any but a set of measure
zero of perturbations.
\bibitem{bit}
The terms ``bits'' and ``qubits'' are used as units
for information content and do not imply a set of two-level systems.
\end{thebibliography}
\end{document}